\documentclass[doublecol]{epl2} 
\usepackage{graphicx}
\usepackage{amsmath,amssymb,bm}
\usepackage{color}

\usepackage{mathptmx}
\usepackage{psfrag}
\usepackage{verbatim}

\newcommand{\beq}{\begin{equation}}
\newcommand{\eeq}{\end{equation}}

\newcommand{\br}{\mathbf{r}}
\newcommand{\bp}{\mathbf{p}}

\newcommand{\bP}{\mathbf{P}}
\newcommand{\bV}{\mathbf{V}}

\title{On the microscopic foundation of dissipative particle dynamics}

\author{A. Eriksson\inst{1,2} \and M. Nilsson Jacobi\inst{2} \and J. Nystr\"{o}m\inst{2} \and K. Tunstr{\o}m\inst{2}}
\shortauthor{A. Eriksson \etal}

\institute{                    
  \inst{1}Department of Physics, University of Gothenburg, SE-41296 G\"oteborg, Sweden\\
  \inst{2} Complex Systems Group, Department of  Energy and Environment, Chalmers University of Technology, SE-41296 G\"oteborg, Sweden
}


\abstract{ 
Mesoscopic particle based fluid models, such as dissipative particle dynamics, are usually assumed to be coarse-grained representations of an underlying microscopic fluid. A fundamental question is whether there exists a map from microscopic particles in these systems to the corresponding coarse-grained particles, such that the coarse-grained system has the same bulk and transport properties as the underlying system. In this letter, we investigate the coarse-graining of microscopic fluids using a Voronoi type projection that has been suggested in several studies. The simulations show that the projection fails in defining coarse-grained particles that have a physically meaningful connection to the microscopic fluid. In particular, the Voronoi projection produces identical coarse-grained equilibrium properties when applied to systems with different microscopic interactions and different bulk properties.
}

\begin{document}
\maketitle


\section{Introduction}

Despite the tremendous computing power available today at the technological high-end of super computing and distributed computing, detailed first-principles molecular simulations are still limited to structures and mechanism on small time and length scales. For instance, molecular dynamics (MD), a well developed framework for molecular simulations, is at present capable of accurate modeling of systems up to typically millions of atoms on a time scale of about $100$ ns~\cite{klein08}. This is enough to simulate small viruses~\cite{freddolino06} or molecular motors~\cite{kitao06}, but to model larger parts of a complex biomembrane or an entire organelle in a biological cell, there is a gap in time and length scale that will not be closed in any near future by increasing the computational resources. This limits our understanding of organization and dynamics at mesoscopic scales--a challenge that applies to biology, soft and hard matter physics~\cite{laughlin00}--- and has resulted in  large efforts invested in developing simulation techniques that allow for exploration of the mesoscopic regime.

In soft matter physics, mesoscopic particle based techniques has gained popularity as tools for simulating complex fluids. Dissipative particle dynamics, smoothed particle dynamics, and a range of descending variations are representative examples. A basic premise of these methods is that there exists a valid mesoscopic particle representation of fluid that in principle can be derived bottom-up, with interactions that give the correct hydrodynamic behavior in the large system limit. The aim of this letter is to reexamine the validity of this premise. We focus on the foundation of the class of models referred to as dissipative particle dynamics. 

Dissipative particle dynamics (DPD) was introduced in the early nineties by Hoogerbrugge and Koelman \cite{hoogerbrugge_koelman92} as a simulation method for complex fluids. The method was motivated partly by the need to counter the problems of broken isotropy and Galilean invariance in lattice gas methods (LGA), partly by the need of a simpler simulation scheme for complex fluids. The remedy was found in the construction of a molecular dynamics (MD) like scheme: A set of `fluid particles' interacting with pairwise (and central) conservative, dissipative and stochastic forces. This construction ensures local conservation of linear and angular momentum, a necessity for obtaining correct hydrodynamic behavior in the macroscopic limit \cite{espanol95, marsh_etal97_FPB}. The original version of DPD was later modified by Espa\~nol and Warren \cite{espanol_warren95} into the form most frequently encountered in the literature today. In this version the dissipative and stochastic forces fulfill a fluctuation-dissipation relation such that the steady state solution of the simulated system (in the continuous time limit) is the Gibbs canonical ensemble. 

With the advantages of being a simple off-lattice and hydrodynamically correct method, DPD is often mentioned as one of the best approaches to model and simulate complex fluids on different time and length scales. As DPD has become a popular mesoscopic simulation tool, it has been applied to an extensive range of systems, from suspensions~\cite{martys05}, polymer systems~\cite{florent08}, phase separations~\cite{groot_madden98}, and membrane and vesicle formation~\cite{shillcock_lipowsky02, yamamoto_etal02}, to modeling of red blood cells \cite{pivkin_karniadakis08} and the life cycle of a minimal protocell \cite{fellerman_etal07}. The span in applications is reflecting the flexibility of DPD, allowing it to be easily adapted to different types of systems, on different time and length scales. But the same flexibility also forces an important question: What is the physical meaning of a DPD particle? 
 
Throughout the literature the DPD particles are described in various ways as ``fluid packets" \cite{koelman_hoogerbrugge92}, ``$\dots$ fluid regions, rather than individual atoms and molecules" \cite{allen06}, and ``$\dots$ not a solvent molecule but a fluid element, which represent clusters of solvent molecules" \cite{noguchi_gompper07}. These are just a few samples, but representative of how DPD is assumed to connect to an underlying microscopic system. In summary, the consensus seems to be that a DPD particle represents some sort of clustered representation of particles in the corresponding microscopic system. How, and more important, if this connection can be formulated is not known. 

Considering this background we believe it is timely and necessary to ask if the concept of mesoscopic fluid particles has a sound physical interpretation. In technical terms the question is: Does there exist a projection that applied to a microscopic fluid results in DPD dynamics on the mesoscopic level? The existence of this type of projection is crucial for particle based methods to have a dynamical interpretation, where the trajectories of the coarse-grained particles have a meaning beyond re-creating certain equilibrium thermodynamical properties of the system. 

This view of DPD as being a coarse-grained representation is often put forward in the literature, usually by referring to the Mori-Zwanzig projection operator framework. There are also several reports on the formal derivation of DPD from projection operators, e.g. the GENERIC framework developed by Espa\~{n}ol~\cite{espanol03}. However, it is harder to find studies that more explicitly state which projection operator to use in the derivation of DPD. The best known attempt to work out a bottom-up derivation of DPD with an explicit projection was done by Flekk{\o}y and Coveney \cite{flekkoy_etal99}. By assuming the DPD particles to be Voronoi cells, they obtain the equations of motion for the mesoscopic system through a formal coarse-graining of the microscopic dynamics. While the resulting equations are DPD-like they do not answer if the actual dynamics of the Voroni cells, as given by the projection, can be approximated by the standard DPD equations. Another study appeared more recently, where a projection operator approach was used to derive the equations of motion for clusters of particles \cite{kinjo07}. In this derivation, however, the result relies on the assumption that the microscopic particles cannot move between clusters, which seems irreconcilable with the fluid character of the microscopic dynamics.

In this letter we investigate using MD simulations how the equilibrium and transport properties of microscopic fluids carry through to the mesoscopic scale. This is done with projections defined through Voronoi tesselation of the microscopic dynamics. In the light of the results, we discuss the implications for mesoscopic particle methods.


\section{Coarse-graining}

To obtain a coarse-grained representation of a microscopic particle system, it is necessary to define a mapping from the microscopic phase space to a coarse-grained level, reducing the number of degrees of freedom. There are many ways to choose a projection. But having in mind the picture of DPD as `fluid particles', a projection defined by Voronoi tesselation of the simulation box is arguably a representation that is natural, as previously pointed out in the literature~\cite{espanol_etal97, flekkoy_etal99}.

\begin{figure}
\centering
\psfrag{x}[][]{$x$ coordinate}
\psfrag{y}[][]{$y$ coordinate}
\includegraphics[width=4.5cm]{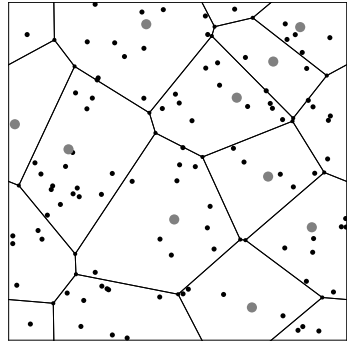}
\caption{\label{fig: voronoi}Periodic Voronoi tesselation in two dimensions defined by coarse-grained particles (gray). Each microscopic particle (black) is assigned to its closest coarse-grained particle and lies within the Voronoi cell of its assigned particle. The Voronoi cells vary both in number of microscopic particles within their borders, and their shape.}
\end{figure}

The procedure works like this: A set of coarse-grained particles, or clusters, are placed in the same simulation box as the microscopic system. The number of clusters depends on the level of coarse-graining.  Each microscopic particle is assigned to the coarse-grained particle closest to itself; this defines a Voronoi tesselation of the simulation box. In Fig.~~\ref{fig: voronoi} this is illustrated for the two dimensional case with periodic boundaries, with coarse-grained particles in gray and microscopic particles in black. Note that the shape of a Voronoi cell can vary as well as the number of microscopic particles within a cell. We denote the average cluster size, i.e. the average number of microscopic particles per cluster, by $N$.

With each microscopic particle assigned to a coarse-grained particle, we define a mapping of the microscopic dynamics to the coarse-grained level in terms of the microscopic variables mass $m_i$, position $\br_i$, and momentum $\bp_i$, where $i=1 \dots n$ denotes microscopic particle $i$. The coarse-grained dynamics can be written as
\begin{eqnarray}
	\label{eq: projection}
	M_k &=& \sum_{i=1}^n\xi_k(\br_i)m_i \Big/ \sum_{i=1}^n\xi_k(\br_i) , \nonumber \\
	\bP_k &=&\sum_{i=1}^n \xi_k(\br_i) \bp_i \Big / \sum_{i=1}^n\xi_k(\br_i) , \nonumber \\
	 \bV_k &=& 	\bP_k/ M_k ,
\end{eqnarray} 
where $M_k$, $\bP_k$, and $\bV_k$  are the mass,  momentum, and velocity of cluster $k$ respectively. In general, $\xi_k(\br_i)$ is a non-negative function which gives the relative contribution of particle $i$ to cluster $k$. This formulation guarantees conservation of momentum and mass. In the Voronoi projection $\xi_k(\br_i)$ is $1$ if microscopic particle $i$ is closest to coarse-grained particle $k$, otherwise $0$. The projected mass of a coarse-grained particle is just the sum of all the microscopic mass within its Voronoi cell, as the projected momentum is the sum of the microscopic momentums within the same cell.  


\section{Simulation set up}
We investigate how the equilibrium and transport properties of a coarse-grained fluid reflect the properties of the underlying system by running simulations along two directions: first, we examine how  the coarse-grained system depend on the average cluster size $N$. This is done using a Lennard-Jones fluid as underlying system, simulated with standard Velocity Verlet MD using the shifted and truncated pairwise Lennard-Jones potential with reduced units
\beq
	U_\text{L-J}(r) = 4 \left[ r^{-12}-r^{-6}- r_c^{-12}+ r_c^{-6}\right] H(r_c - r),
\eeq
where $r$ is the distance between particle pairs, $r_c$ is cutoff radius, and $H(x)$ is the Heaviside step function which is one for $x>0$ and zero elsewhere. In the Lennard-Jones simulations, $r_c=3$.

Second, we apply an identical projection, average cluster size $N=10$, to microscopic systems with different interaction potentials to find out how sensitive the projection is to the details of the underlying system. In addition to the Lennard-Jones fluid, we simulate fluids with quadratic and linear potentials 
\begin{eqnarray}
	U_\text{Q}(r)&=&0.5 \, a \, r_c (1-r/r_c)^2 H(r_c - r) \quad \text{and}\\
	U_\text{L}(r)&=&a \, r_c (1-r/r_c) H(r_c - r).
\end{eqnarray}
The pre-factor $a$ defines the magnitude of the force. Two quadratic potential fluids are simulated: $U_{\text{Q}_1}$ with $a=25$ and $r_c=1$ and $U_{\text{Q}_2}$ with $a=100$ and $r_c=1$. The linear potential fluid is simulated with $a=25$ and $r_c = 1$.

All microscopic systems are simulated in the microcanonical ensemble (constant volume, energy and number of particles) using periodic boundary conditions. The parameter settings are time step $\tau = 0.005$, particle density $\rho=0.776$ and temperature $T=0.861$ (all in reduced units), chosen so that the Lennard-Jones system is in the fluid regime. This corresponds to the situation when DPD is usually assumed to represent the mesoscopic behavior. All projections are done with $200$ clusters, with box size and number of microscopic particles varied accordingly to obtain the correct average cluster size $N$. The cluster positions are initially random, and then updated using the velocity in Eq.~(\ref{eq: projection}).
The simulations are run to equilibrium or steady state equilibrium before any measurements are obtained. All results are averaged over four measurements from independent simulations. Both equilibrium and transport properties of the coarse-grained system are measured directly from the resulting cluster coordinates and momentums.


\section{Simulation results}
For a particle system interacting with pairwise and central forces there is a one-to-one correspondence between the radial distribution function (RDF) and the pairwise potential~\cite{henderson74}. This means that the equilibrium properties of a particle system at given density and temperature are uniquely defined by the RDF. In order to check that the projection conserves the equilibrium properties, we compare the coarse-grained RDFs and the bulk moduli for the different microscopic and coarse-grained systems. We also measure the mass distribution of clusters, and relate this to the standard DPD model. How the transport properties of the microscopic system translate to the projected system is examined by comparing the diffusion and viscosity at each level.

\subsection{Radial distribution functions}

\begin{figure}
\centering
\psfrag{RDF}[][]{RDF}
\psfrag{Distance}[][]{$r$}
\includegraphics[width=8cm]{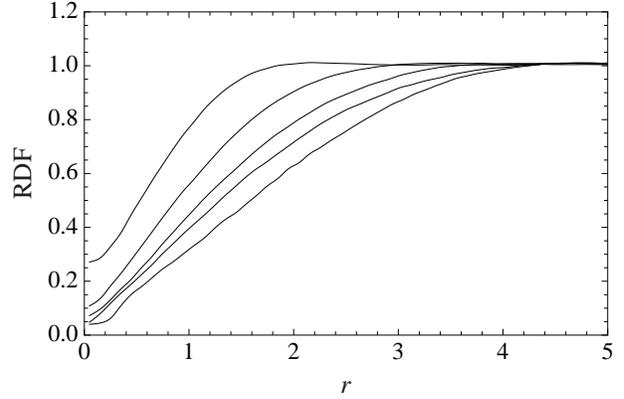}
\caption{\label{fig: rdfClusters}Radial distribution functions of coarse-grained systems. The curves correspond to systems with a different average number of microscopic particles per cluster (from left to right: 3, 10, 20, 30, 40). The Lennard-Jones fluid is the underlying system in each case.}
\end{figure}

The RDFs of a coarse-grained Lennard-Jones fluid, using average cluster sizes from $N=3$ to $N=40$, are presented in Fig.~~\ref{fig: rdfClusters}. The RDFs are characterized by having a slow, almost linear, climb from zero outwards, before smoothing off to one, the slope depending on the size of the cluster. In terms of interactions, this kind of behavior is typically of `soft' interactions, such as the interactions typically used in standard DPD, and can be attributed to an excluded volume effect of the Voronoi cells.

\begin{figure}
\centering
\psfrag{RDF}[][]{RDF}
\psfrag{Distance}[][]{$r$}
\includegraphics[width=8cm]{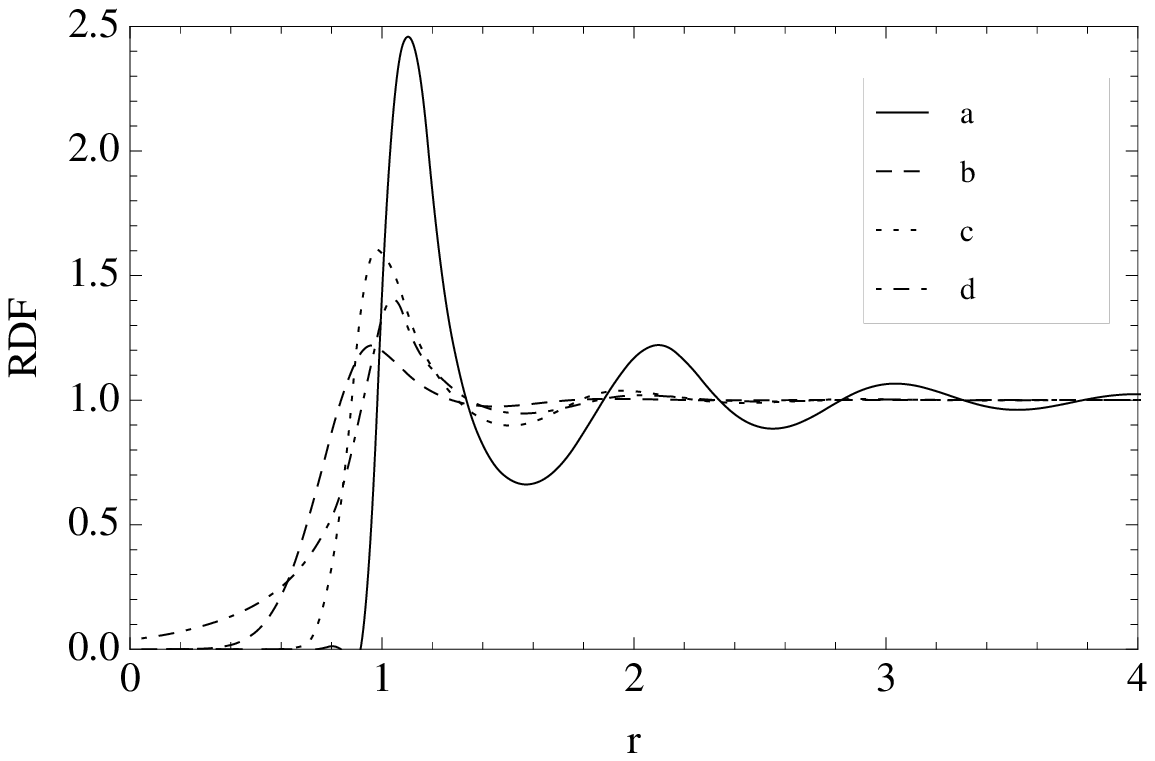}
\includegraphics[width=8cm]{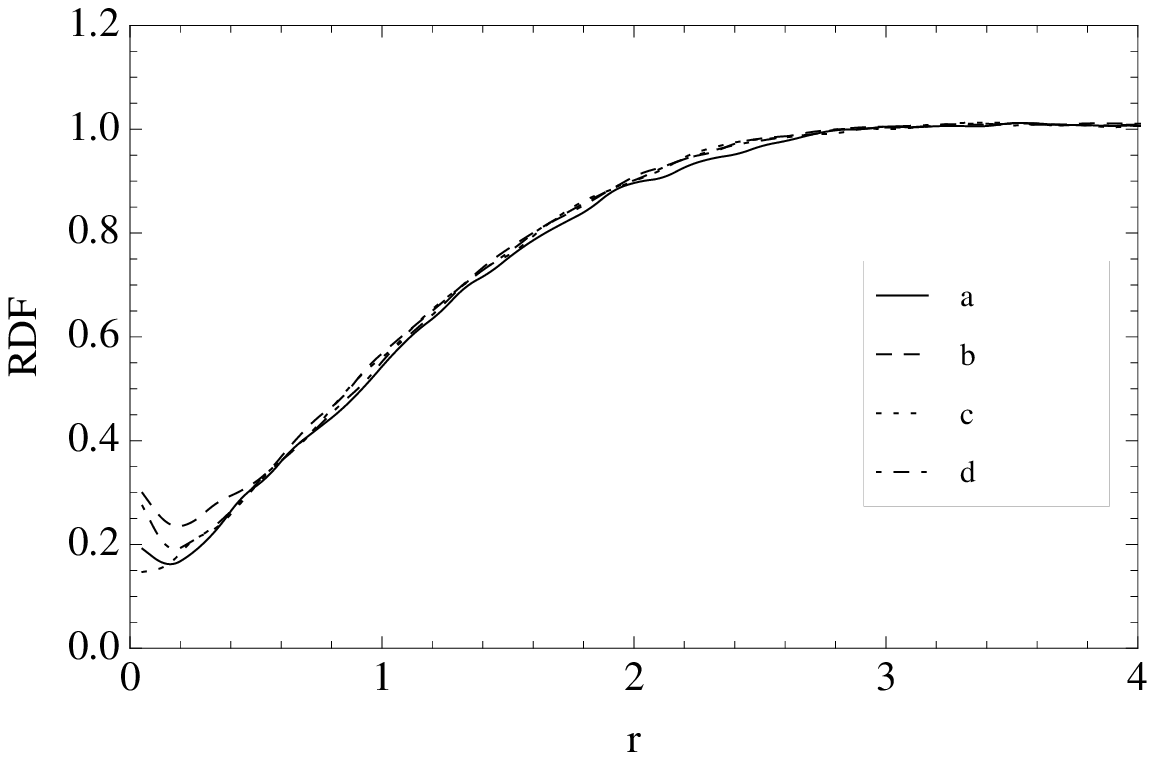}
\psfrag{RDF}[][]{RDF}
\psfrag{Distance}[][]{$r$}
\caption{\label{fig: rdfSub} Upper figure: Radial distribution functions of different microscopic systems. a) Lennard-Jones potential $U_{\text{L-J}}$; b) Quadratic potential $U_{\text{Q}_1}$; c) Quadratic potential $U_{\text{Q}_2}$; and d) Linear potential $U_{\text{L}}$.  Lower figure: Radial distribution functions of corresponding coarse-grained systems, all with average cluster size $N=10$.}
\end{figure}

Close to zero, the RDFs flatten out. This behavior can be attributed to individual clusters having zero mass, and this region must therefore be treated with care. This has been described previously in ref.~\cite{espanol_etal97}. The RDF for the underlying Lennard-Jones fluid is given by curve (a) in Fig.~~\ref{fig: rdfSub}.

To explore the sensitivity of the projection to the underlying system, we applied the projection with average cluster size $N=10$ to different microscopic systems.	The RDFs for the microscopic systems and the corresponding cluster systems are all plotted in Fig.~~\ref{fig: rdfSub}. While the microscopic systems have clearly distinguishable RDFs, the corresponding cluster RDFs are almost identical. The discrepancy in the cluster RDFs for small $r$ is an artifact of the projection as discussed above. 
\subsection{Bulk modulus}
For the different microscopic systems, we estimated the bulk moduli $B$  (or the inverse compressibility) using the expression 
\beq
	B=\rho \left( \frac{\partial p}{\partial \rho} \right)_T, 
\eeq
where $p$ is the pressure of the system, and the partial derivative is evaluated at constant temperature $T$. To obtain a numerical estimate for the derivative, the microscopic systems were simulated with densities slightly higher and lower than $\rho=0.776$. For each density the pressure was measured using the virial expansion, and the derivative was  calculated as pressure difference over density difference. The bulk moduli for the corresponding clusters were obtained in the same way. To be able to apply the virial expansion in this case, we derived the effective pairwise interactions of the clusters, estimated from the RDFs using a inverse Monte Carlo technique~\cite{henderson74,lyubartsev03}. The resulting bulk moduli are listed in Table~\ref{table: bulk}. The microscopic systems have different moduli, while the coarse-grained systems are almost identical and seemingly independent of the underlying system. 
\begin{table}
\caption{\label{table: bulk}
Bulk modulus for the different microscopic systems and corresponding coarse-grained systems. The numbers in parentheses are the uncertainty of the last digit.}
\smallskip
\centering	
\begin{tabular}{l@{\extracolsep{1em}}  c c}
	Potential 			& $B_\text{M}$ 	& $B_\text{C}$    \\
	\hline
	$U_\text{L-J}$		& $8.1(1)$ 		& $0.25(1)$ \\
	$U_{\text{Q}_1}$ 	& $5.4(2)$ 		& $0.25(1)$ \\
	$U_{\text{Q}_2}$		& $2.6(1)$ 		& $0.25(1)$ \\
	$U_\text{L}$		& $15.7(1)$ 		& $0.25(1)$ \\
\end{tabular}
\end{table}
\subsection{Diffusion}

\begin{figure}
\centering
\psfrag{N}[][]{Average cluster size $N$}
\psfrag{DN}[][]{Diffusion $D$}
\includegraphics[width=8cm]{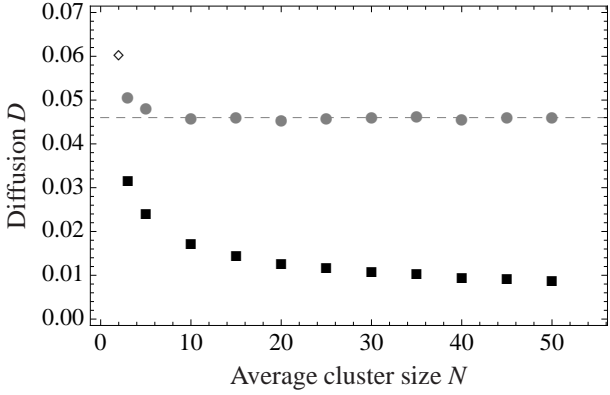}
\caption{\label{fig: diffusion} Diffusion of clusters plotted against average cluster size $N$ (black squares). The diffusion of each cluster divided by $N^{0.43}$ is approximately constant (gray circles and dashed line), indicating a scaling law of the diffusion. The diffusion of the underlying Lennard-Jones fluid is shown as a reference (diamond).} 
\end{figure}

The diffusion coefficients were calculated from the mean square displacement of each system. Projections of a Lennard-Jones fluid, using different average cluster sizes $N$, produce the coefficients plotted in Fig.~~\ref{fig: diffusion} (black squares), with standard deviation on the scale of the marker size. The diffusion coefficient of the underlying Lennard-Jones fluid is given by the diamond marker. We observe that the diffusion falls off with increasing $N$. The values we get by dividing each diffusion coefficient with the empirical factor $N^{-0.43}$ are nearly constant (gray circles and dashed line), and suggest an approximate power-law relation between cluster size and cluster diffusion. 

The diffusion coefficients for the different microscopic systems and corresponding cluster systems (all with $N=10$) are listed in Table~\ref{table: diffusionSub}, together with the ratio between the coefficients. The variation in the ratios implies that the diffusion not only scales with $N$, but also with the underlying dynamics.
\begin{table}
\caption{\label{table: diffusionSub}
Diffusion coefficients for the different microscopic systems and corresponding coarse-grained systems. The last row shows the ratio between the coefficients. The numbers in parentheses are the uncertainty of the last digit.}
\smallskip
\centering	
\begin{tabular}{l@{\extracolsep{1em}}  c c c}
	Potential &			$D_\text{M}$ & 	$D_\text{C}$ &	$D_\text{M}/D_\text{C}$  \\
	\hline
		$U_\text{L-J}$ &		$0.0642(3)$ & 	$0.0168(4)$ &	$3.82$\\
		$U_{\text{Q}_1}$ & 	$0.527(4)$ & 	$0.083(2)$ & 	$6.35$\\	
		$U_{\text{Q}_2}$ & 	$0.248(1)$ & 	$0.058(2)$ &		$4.28$\\
	   	$U_\text{L}$	&	$1.47(3)$ & 		$0.200(3)$ &		$7.35$\\
\end{tabular}
\end{table}
\subsection{Viscosity}

The shear viscosity was measured using a Poiseuille flow method~\cite{backer_etal05}. The external force used to drive the system to a steady state flow was applied to the microscopic system, not affecting any steps in the coarse-graining. Measurements were done for the cluster systems for average cluster sizes $N=3,20,30$ and $40$. As a check on the influence of the box size on the simulations, we also measured the viscosity of the Lennard-Jones fluid in each case. The resulting values are listed in Table~\ref{table: viscosity}. The viscosity is consistently higher (appr. $5$--$10\%$) for the clusters, but does not vary significantly with size (at least not more than the microscopic system does).
\begin{table}	
\caption{\label{table: viscosity}Viscosity values $\mu_{C}$ for cluster systems with different average cluster size $N$, projected from the same  Lennard-Jones system. The Lennard-Jones viscosity values $\mu_{L-J}$ are shown as a check for box size sensitivity. The numbers in parentheses are the uncertainty of the last digit. The last column lists standard deviation in cluster size.}
\smallskip
\centering
\begin{tabular}{c@{\extracolsep{1em}}ccc}
	$N$	&   $\mu_{C}$	& $\mu_{L-J}$ 	& $\sigma_N$\\
	\hline
	3		 & $1.94(8)$	& 	$1.80(5)$ &	$1.24$		\\
	20		 & $1.96(4)$	& 	$1.85(3)$ & 	$5.79$ 	\\
	30		 & $1.98(4)$	& 	$1.83(3)$ &	$8.50$		\\
	40		& $2.02(4)$ 	& 	$1.87(1)$ & 	$10.39$	\\
\end{tabular}
\end{table}
\subsection{Mass distribution}

\begin{figure}
\centering
\psfrag{Size}[][]{Average cluster size $N$}
\psfrag{Distribution}[][]{Distribution}
\includegraphics[width=8cm]{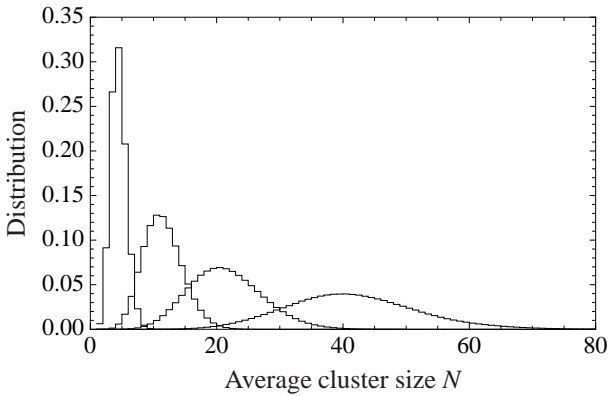}
\caption{\label{fig: histogram}Histograms showing the size distribution for cluster systems with different average cluster size $N$. From left to right are the histograms for systems with $N=3, 10,20$ and $40$.}
\end{figure}

Because the projection allows microscopic particles to move between the clusters, the sizes of the individual clusters fluctuated around the mean value $N$. Fig.~~\ref{fig: histogram} contains mass distributions measured in systems with average cluster sizes from $N=3$ to $40$, all projected from the same Lennard-Jones fluid. The distribution becomes broader with increasing $N$. In fact,  due to the fluctuating volumes in the Voronoi tesselation, the standard deviation scales proportionally to $N$ (Table \ref{table: viscosity}). This in contrast to the standard DPD set up, where all particles have constant, and usually identical, mass.


\section{Discussion}

In this letter we have investigated the feasability of a microscopic foundation of coarse-grained particle dynamics. Specifically, we try to answer the following question: does a Voronoi projection produce a coarse-graining that can be interpreted as mesoscopic particle dynamics with pairwise interactions? The motivation behind the question is that a class of simulation techniques, related to the so called dissipative particle dynamics (DPD) method, is based on the assumption that this, or perhaps some other, projection actually produce well defined particle dynamics on a mesoscale. 

The results that we find can be summarized as follows: The self-diffusion rate of the cluster centers obey a scaling law (empirically observed); The viscosity of the projected system is similar to the underlying system, and does not depend on the number of particles; The standard deviation of average cluster mass scales approximately linearly with the cluster mass, and implies that the clusters do not get a more well defined mass as the system grows; The cluster-cluster radial distribution function is very similar for different microscopic systems, also reflected in the bulk modulus (and compressibility) of the system. 

For all microscopic systems, the bulk modulus is much higher than in the corresponding coarse-grained system. This difference can be partly attributed to that we do not take the internal pressure of the coarse-grained particles into account. A more serious difficulty is that the coarse-grained bulk modulus is approximately the same for all systems, whereas at the microscopic level the  bulk modulus varies significantly. It is difficult to see how the coarse-grained dynamics could simultaneously have an interaction that is consistent with the microscopic dynamics and recover the correct bulk modulus in the macroscopic limit. This is especially worrying when we consider that the standard approach to choosing  the DPD parameters is to tune the interaction strength so that the macroscopic  bulk modulus (or equivalently, the compressibility) is correct.

\begin{figure}
\centering
\psfrag{Distance}[][]{$r$}
\psfrag{RDF}[][]{RDF}
\includegraphics[width=8cm]{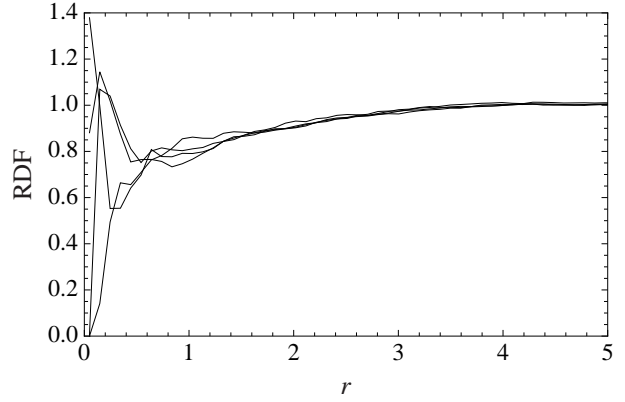}
\caption{\label{fig: softcluster}Radial distribution functions for different realizations of the same coarse-grained system. The coarse-grained dynamics is resulting from a projection of the microscopic dynamics through a weighing function (see text). The curves have not reached equilibrium after half a million time steps.}
\end{figure}

A potential source of the problems with the Voronoi projection is that the momenta of the cluster centers changes discontinuously when a microscopic particle moves from one cluster to another. In order to avoid this situation, we lifted the restriction that a microscopic particle only belongs to its closest neighboring coarse-grained particle. This was done by replacing $\xi_k(\br_i)$ in Eq.~(\ref{eq: projection})  by a weighing function which decreases exponentially with the distance between the cluster center and the microscopic particle. See ref.~\cite{flekkoy_etal99} for details. A microscopic particle is then distributed among several coarse-grained particles, contributing to the mass and momentum of each coarse-grained particle according to the weighing functions. This results in a soft clustering rather than the sharp Voronoi partition used above. In Fig.~~\ref{fig: softcluster} we show RDFs from the same coarse-grained  system, but with different initial conditions. After half a million time steps, the curves have not converged. In contrast, the RDF of the microscopic particles converge in less than a few thousand time steps after an initial temperature equilibration period of ten thousand steps. The slow convergence is caused by coarse-grained particles close together having strongly correlated dynamics, staying together for long periods. While not completely understood, it seems to be connected to the mass variations in clusters.
Therefore, this projection is not a practical choice for coarse-graining.

Another type of projection that fits the fluid particle view, is to have the positions of the coarse-grained particles defined through K-mean clustering of the underlying particles \cite{eriksson08a}. The main motivation behind such a projection is that it more directly reflects the spatial density variations in the microscopic system, whereas the Voronoi projection does not have such a direct connection to the physical distribution of the underlying particles. However, as we have shown in earlier work~\cite{eriksson08a}, using this projection is problematic as it results in a coarse-grained dynamics which suffers from discontinuous trajectories, and it does not conserve the  momentum of the particles.

Our main conclusion is that the Voronoi projection has severe shortcomings as a means to define effective particles at a coarse-grained level from the dynamics of the microscopic particles. However, alternative projections fail even more drastically. It is therefore an open question, does there exist a projection from microscopic dynamic to the mesoscopic level such that the mesoscopic level has a particle dynamic?
We view this as a serious challenge to the community working with mesoscopic particle based simulation methods. If DPD and related methods are to become well established the question about which projection does define the coarse-graining is critical. 

There are two ways to argue around the problems pointed out in this letter. First, it is  still possible that the Voronoi projection does produce a useful coarse-graining, however not in terms of a mesoscale point particle dynamics. The meso-particles can instead be interpreted as nodes in a dynamic mesh representing local mechanical variables. Smoothed dissipative particle dynamics~\cite{espanol97} may be viewed as an attempt in this direction even though the model still uses the particle interpretation at the mesoscale (which is central if local momentum conservation is to be respected). It should be remembered that with this interpretation it is no longer possible to measure observables such as compressibility and diffusion directly from the trajectories of the meso-particles or rather the nodes in the mesh. As a consequence most of the machinery usually applied in connection with DPD must then be abandoned. Alternatively, as in \cite{flekkoy_etal99} one abandons the idea of point particle dynamics at the mesoscopic level and simulates the motion of the Voroni regions.

The second remedy is to view DPD as a pure top-down modeling approach and simply ignore that there is no clear connection to the microscopic dynamics. Rather, one then treats each mesoscopic particle as a thermodynamic system, as in the GENERIC framework \cite{espanol98,espanol03}, in which the equation of state can be directly specified \cite{pagonabarraga_frenkel01}. With this perspective it is natural to tune the parameters in the model to fit with macroscopic observables. However, we must stress that in this case it is unclear if the DPD model has any value as a dynamic model outside equilibrium. On the other hand, the lack of a clear connection between the scales is not unique. For example the classical measurement problem shows that the connection between quantum mechanics and molecular dynamics is not exactly crystal clear. However, we think that it is unlikely that the relation between MD and mesoscale simulation methods should hide similar level of complexity, especially since the two levels essentially deals with the same type of representation of the respective system.
\acknowledgments
This work was funded (in part) by the EU integrated project FP6-IST-FET PACE, by EMBIO, a European Project in the EU FP6 NEST Initiative, and by the Research Council of Norway. 
\bibliography{arXiv6}
\end{document}